\input{psfig}

\newcommand{\vk}{\mbox {\boldmath $k$\unboldmath}}
\newcommand{\vq}{\mbox {\boldmath $q$\unboldmath}}
\def\nn{\nonumber}
\documentclass[epj,referee]{svjour}
\begin{document}

\title{Quark Model Study of The $\eta$ Photoproduction:}
\subtitle{Evidence for a New $S_{11}$ Resonance?\\}
\author{Bijan Saghai\inst{1} and Zhenping Li\inst{2}.
}                     
% Do not remove
%
%
\institute{$^1$ Service de Physique Nucl\' eaire, DAPNIA - CEA/Saclay,
F-91191 Gif-sur-Yvette Cedex, France \\
$^2$ Physics Department, Peking University,
Beijing 100871, P.R. China}
\date{Received: date / Revised version: date}
% The correct dates will be entered by Springer
%
\abstract{
An extensive and systematic study of the recent $\eta$ photoproduction
data up to 1.2 GeV is presented within a chiral constituent quark model. 
A model embodying all known nucleonic resonances
shows clear need for a yet undiscovered third $S_{11}$ resonance in the 
second resonance region, for which we determine the mass (1.729 GeV) and 
the total width (183 MeV). 
Furthermore, we extract the configuration mixing angles, an 
important property of the quark-quark interaction in the quark model,
for the resonances $ S_{11}(1535)$ and $S_{11}(1650)$, as well as for
the resonances
$D_{13}(1520)$ and $D_{13}(1700)$. Our results agree well with the
quark model predictions. 
In addition, the partial $\eta N$ decay widths and/or the photo-excitation 
helicity amplitudes for the nucleonic resonances $S_{11}(1535)$, 
$S_{11}(1650)$, $P_{11}(1710)$, 
$P_{13}(1720)$, $D_{13}(1520)$, $D_{13}(1700)$, $D_{15}(1675)$, and 
$F_{15}(1680)$ are also obtained in this approach.
 % end of PACS codes
} %end of abstract

\maketitle
\section{Introduction}
\label{sec:Intro}
%%%
Investigation of the $\eta$-meson production {\it via} electromagnetic probes
offers access to several exciting topics in hadrons spectroscopy. 

One prominent example is the search for missing resonances. 
Several such baryons have been
predicted by different QCD inspired approaches and constitute an
strong test of these formalisms
%%%
\footnote{See, e.g., review 
papers~\cite{Rev_CR-00,Rev_BIL-00,Rev_GR-96,Rev_An-93}, and
references therein.}.
Electromagnetic production of such resonances, if they exist, are looked 
for in various mesons production processes.
To our knowledge, the most extensive theoretical results in the quark 
model approach have been reported in Ref.~\cite{Cap92}, 
where the authors, within a relativized pair-creation ($^3P_0$) model, have 
investigated the quasi-two-body decays of baryons and have proceeded to make
comparisons with the available results from partial-wave 
analysis~\cite{PDG,MS-92}. 

Another example, more specific to the $\eta N$ final state, is the 
enhancement~\cite{Dyt00,Nac00,Sva1,Kai95,Deu95} 
of the resonance $S_{11}(1535)$ decaying into the $\eta N$ and the 
suppression of another S-wave resonance $S_{11}(1650)$ in the same channel, 
which provide us with direct insights into the configuration mixings of the
quark model states. A recent work~\cite{GR-96}, embodying the fine structure 
interaction between constituent quarks, has provided a qualitative description 
of the suppressed decay of the $S_{11}(1650) \to N \eta$ compared to the large 
branching ratio for the $S_{11}(1535) \to N \eta$ decay, though the 
electromagnetic couplings of the resonance $S_{11}(1535)$ remain to be 
evaluated in this approach. It has also been suggested~\cite{zr96} that 
quasi-bound $K\Lambda$ or $K\Sigma$ states might be an answer to this puzzle. 

Moreover, the properties of the decay of baryon resonances into $\gamma N$ 
and/or meson-nucleon are intimately related to their internal 
structure~\cite{Cap92,KI-80,CI-86,LC90,BIL}.
Extensive recent experimental efforts on the $\eta$ 
photo-~\cite{BB,Mainz,elsa,graal,Graal2000,Fred}
and electro-production~\cite{Armstrong,CLAS_e} 
are opening a new era in this topic.
The process $\gamma p \to \eta p$, with real or virtual photons, has been 
proven~\cite{LS-1,Tiator-99,Kno95,RPI95a,RPI95b,RPI97,Nimai-98,Dav00}
to be very attractive in the extraction of the photo-excitation 
amplitudes of the $S_{11}(1535)$ and/or $D_{13}(1520)$ resonances. 
Perhaps more importantly, the data make it possible to improve the accuracy 
in the determination of the $N^* \to \eta N$ branching ratios.

All these features can be studied through the $\eta$-meson photoproduction.
At the present time, near threshold region has been studied extensively
{\it via} a variety of formalisms, such as 
effective Lagrangian
approaches~\cite{Kno95,RPI95a,RPI95b,RPI97,Nimai-98,Dav00,Lin94,Tia94},
generalized Lee model~\cite{Den98},
coupled channel calculations~\cite{FM,Gre99,Deu97,CB91},
chiral meson-baryon Lagrangian theory~\cite{kai97,car00,Bor00}, and 
constituent
quark formalism~\cite{LS-1,zpl95,zpl97,ZSL,Sta95}.

These efforts have considerably improved our understanding of the
underlying elementary reaction mechanism at low energy.
Here, the most quantitative phenomenological investigations concern 
the first resonance region, where the differential and total
cross-section data obtained at 
Mainz~\cite{Mainz}, for $E_{\gamma}^{lab} <$ 0.8 GeV have been extensively
exploited. Some of those works include also target polarization asymmetry form
ELSA~\cite{elsa}, and/or polarized beam asymmetry from Graal~\cite{graal}.
The main finding on the reaction mechanism is the dominance of the
$S_{11}(1535)$ resonance and a small contribution from the $D_{13}(1520)$
resonance. Moreover, these studied have concentrated on 
putting constraints on the $S_{11}(1535)$, and to a less
extent on the $D_{13}(1520)$ resonances parameters.

Very recent differential and total cross section data from 
Graal~\cite{Graal2000} cover
both first and second resonance regions and constitute a real break through 
in this field.  

The focus of this paper is to study all the recent $\gamma p \rightarrow \eta p$
data for $E_{\gamma}^{lab} <$ 1.2 GeV ($W \equiv E_{total}^{cm} <$ 1.75 GeV)  
within a chiral constituent quark formalism based on the $SU(6)\otimes O(3)$ symmetry. 
The advantage of the quark model for the meson photoproduction is the ability 
to relate the photoproduction data directly to the internal structure of the
baryon resonances. 
To go beyond the exact $SU(6)\otimes O(3)$ symmetry, we introduce
symmetry breaking coefficients $C_{N^*}$ as in our earlier 
publication~\cite{LS-1}. 
We further
show how these coefficients are related to the configuration mixing
angles generated by the gluon exchange interactions in the quark 
model~\cite{IK_77,IKK_78}. 
Indeed, our 
extracted mixing angles for the $S$ and $D$ wave resonances in the second resonance
region show very good agreement with the quark model predictions~\cite{IK_77}.

Our main finding in the present work is {\it the need for a third $S_{11}$ resonance
in the second resonance region}, as seemingly dictated by the Graal cross-section data~\cite{Graal2000}
above  $E_{\gamma}^{lab} \approx$ 1.0 GeV. Such a resonance has been 
predicted by the authors of Ref.~\cite{zr96}. Our extracted values for the
mass and width of this resonance agree very well with those put forward
in that paper. If this is confirmed by more accurate and/or higher
energy data, then one possible conclusion would be that
this resonance can not be accommodated by the constituent
quark model, indicating a molecular type of structure~\cite{zr96}. 

In addition,  we present a framework for extracting the $\eta N$ branching
ratios from the data beyond the resonances in the threshold region of the
$\eta$ photoproduction.

This paper is organized as following.
In the next Section, we summarize the theoretical basis of our work,
introduce the configuration mixing angles and relate them to the 
$SU(6)\otimes O(3)$ symmetry breaking coefficients. We present also
expressions for photo-excitation helicity amplitudes and strong decay
widths. Section 3 is devoted to our numerical results. We start with
comparisons between our results and differential cross-section data.
Results for mixing angles are given and the need for a new resonance is
underlined. Then we proceed to comparisons with total cross-section
and polarization observables and
show the role played by the third $S_{11}$ resonance. The obtained model is
then used to extract the helicity amplitudes and strong decay widths.
In Section 4 we summarize our work and end it with some concluding remarks.

%%%%%%%%%%%

\section{Theoretical frame}
\label{sec:Theory}

The starting point of the meson photoproduction in the chiral quark model is the low 
energy QCD Lagrangian~\cite{MANOHAR}
\begin{equation}\label{eq:Lagrangian}
{\cal L}={\bar \psi} \left [ \gamma_{\mu} (i\partial^{\mu}+ V^\mu+\gamma_5
A^\mu)-m\right ] \psi + \dots
\end{equation}
where $\psi$ is the quark field  in the $SU(3)$ symmetry,
$ V^\mu=(\xi^\dagger\partial_\mu\xi+\xi\partial_\mu\xi^\dagger)/2$ 
and 
$A^\mu=i(\xi^\dagger \partial_{\mu} \xi -\xi\partial_{\mu} \xi^\dagger)/2$ 
are the vector and axial currents, respectively, with $\xi=e^{i \Pi f}$; 
$f$ is a decay constant and the field $\Pi$ is a $3\otimes 3$ matrix
\begin{equation}\label{eq:Pi}
\Pi=\left| \begin{array}{ccc} \frac 1{\sqrt {2}} \pi^\circ+\frac 1{\sqrt{6}}\eta 
& \pi^+ & K^+ \\ \pi^- & -\frac 1{\sqrt {2}}\pi^\circ+\frac 1{\sqrt {6}}\eta & 
K^\circ \\ K^- & \bar {K}^\circ &-\sqrt{\frac 23}\eta \end{array}\right|,
\end{equation}
in which the pseudoscalar mesons, $\pi$, $K$, and $\eta$, are treated
as Goldstone bosons so that the Lagrangian in Eq.~(\ref{eq:Lagrangian}) 
is invariant under the chiral transformation.  
Therefore, there are four components for the photoproduction of
pseudoscalar mesons based on the QCD Lagrangian,
\begin{eqnarray}\label{eq:Mfi}
{\cal M}_{fi}&=&\langle N_f| H_{m,e}|N_i \rangle + \nn \\
&&\sum_j\bigg \{ \frac {\langle N_f|H_m |N_j\rangle 
\langle N_j |H_{e}|N_i\rangle }{E_i+\omega-E_j}+ \nn \\
&& \frac {\langle N_f|H_{e}|N_j\rangle \langle N_j|H_m
|N_i\rangle }{E_i-\omega_m-E_j}\bigg \}+{\cal M}_T
\end{eqnarray}
where $N_i(N_f)$ is the initial (final) state of the nucleon, and 
$\omega (\omega_{m})$ represents the energy of incoming (outgoing) 
photons (mesons).  

The first term in Eq.~(\ref{eq:Mfi}) is a seagull term. It is generated by the gauge 
transformation of the axial vector $A_{\mu}$ in the QCD Lagrangian.
This term, being proportional to the electric charge of the outgoing mesons, does 
not contribute to the production of the $\eta$-meson.
The second and third terms correspond to the {\it s-} and {\it u-}channels,
respectively. 
The last term is the {\it t-}channel contribution and contains two parts: 
{\it i)} charged meson exchanges which are proportional to the charge of outgoing 
mesons and thus do not contribute to the process $\gamma N\to \eta N$;  
{\it ii)} $\rho$ and $\omega$ exchange in the $\eta$ production which are 
excluded here due to the duality hypothesis~\cite{Collins,ST}.
We will come back to this point in Section~4.

The pseudovector and electromagnetic couplings at the tree level are given respectively
by the following standard expressions:
\begin{eqnarray}
H_m~&=~&\sum_j \frac 1{f_m} {\bar \psi}_j\gamma_\mu^j\gamma_5^j \psi_j
\partial^{\mu}\phi_m,\label{eq:Hm} \\
H_e~&=~&-\sum_j e_j \gamma^j_\mu A^\mu ({\bf k}, {\bf r}).\label{eq:He}
\end{eqnarray}

Because the baryon resonances in the {\it s-} and {\it u-}channels are treated as three
quark systems, the separation of the center of mass motion from the internal motions in 
the transition operators is crucial.  
Thus, we use a well established approach~\cite{zpli93-94} to evaluate  the contributions 
from resonances in the {\it s-} and {\it u-}channels.

\subsection{Configuration Mixing}
\label{sec:Theory1}
The general framework for the meson photoproduction, in particular, for
the $\eta$ case, has been given in Refs~\cite{zpl95,zpl97}. 
The transition matrix elements based on the low energy QCD
Lagrangian include the {\it s-} and {\it u-}channel contributions
\begin{equation}\label{eq:Msu}
{\cal M}_{if}={\cal M}_s+{\cal M}_{u}.
\end{equation}
The {\it u-}channel contributions are divided into the nucleon Born
term and the contributions from the excited resonances.  The matrix 
elements for the nucleon Born term is given explicitly, while the 
contributions from the excited resonances above 2 GeV for a given parity 
are assumed to be degenerate so that their contributions could be 
written 
in a compact form~\cite{zpl95}.

The contributions from  the {\it s-}channel resonances can be written as
\begin{eqnarray}\label{eq:MR}
{\mathcal M}_{N^*}=\frac {2M_{N^*}}{s-M_{N^*}(M_{N^*}-i\Gamma(q))}
e^{-\frac {{k}^2+{q}^2}{6\alpha^2_{ho}}}{\mathcal A}_{N^*},
\end{eqnarray}
where  $k=|\vk|$ and $q=|\vq|$ represent the momenta of the incoming photon 
and the outgoing meson respectively, $\sqrt {s}$ is the total energy of 
the system, $e^{- {({k}^2+{q}^2)}/{6\alpha^2_{ho}}}$ is a form factor 
in the harmonic oscillator basis with the parameter $\alpha^2_{ho}$ 
related to the harmonic oscillator strength in the wave-function, 
and $M_{N^*}$ and $\Gamma(q)$ are the mass and the total width of 
the resonance, respectively.  The amplitudes ${\mathcal A}_{N^*}$ 
are divided into two parts~\cite{zpl95,zpl97}: the contribution 
from each resonance below 2 GeV, the transition amplitudes of which 
have been translated into the standard CGLN amplitudes in the harmonic 
oscillator basis, and the contributions from the resonances above 2 GeV
treated as degenerate, since little experimental information is available
on those resonances.

The contributions from each resonance to the $\eta$
photoproduction is determined by introducing~\cite{LS-1} a new set of 
parameters $C_{{N^*}}$, and the following substitution rule for the 
amplitudes ${\mathcal A}_{{N^*}}$:
\begin{eqnarray}\label{eq:AR}
{\mathcal A}_{N^*} \to C_{N^*} {\mathcal A}_{N^*} ,
\end{eqnarray}
so that 
\begin{eqnarray}\label{MRexp}{\mathcal M}_{N^*}^{exp} = C^2_{N^*}
 {\mathcal M}_{N^*}^{qm} ,
\end{eqnarray}
where ${\mathcal M}_{N^*}^{exp}$ is the experimental value of 
the observable, and ${\mathcal M}_{N^*}^{qm}$ is calculated in the 
quark model~\cite{zpl97}. 
The $SU(6)\otimes O(3)$ symmetry predicts
$C_{N^*}$~=~0.0 for ${S_{11}(1650)} $, ${D_{13}(1700)}$, and 
${D_{15}(1675)} $ resonances, and $C_{N^*}$~=~1.0 for other
resonances in Table~1.  
Thus, the coefficients $C_{{N^*}}$ measure the discrepancies between 
the theoretical results and the experimental data and show the extent 
to which the $SU(6)\otimes O(3)$ symmetry is broken in the process 
investigated here. 

%
%%%%%%%%%%%%%%%%%%%%%%%%%%%%  TABLE I
%   
\begin{table}\label{tab:Res}
\caption{Resonances included in our study with their 
assignments in $SU(6)\otimes O(3)$ configurations, masses, 
and widths. The mass and width of the $S_{11}(1535)$ resonance are
left as adjustable parameters (see Table~4).}
\label{assign}
\begin{center}
\begin{tabular}{llccccc}  
States & & $SU(6)\otimes O(3)$& & Mass & & Width   \\  
 & & & &  (GeV) & &  (GeV)  \\  \hline        
$S_{11}(1535)$&&$N(^2P_M)_{\frac 12^-}$&& && \\[1ex] 
$S_{11}(1650)$&&$N(^4P_M)_{\frac 12^-}$&&1.650&&0.150 \\[1ex]    
$D_{13}(1520)$&&$N(^2P_M)_{\frac 32^-}$&&1.520&&0.130\\[1ex]    
$D_{13}(1700)$&&$N(^4P_M)_{\frac 32^-}$&&1.700&&0.150\\[1ex]
$D_{15}(1675)$&&$N(^4P_M)_{\frac 52^-}$&&1.675&&0.150\\[1ex]
$P_{13}(1720)$&&$N(^2D_S)_{\frac 32^+}$&&1.720&&0.150\\[1ex]    
$F_{15}(1680)$&&$N(^2D_S)_{\frac 52^+}$&&1.680&&0.130\\[1ex]    
$P_{11}(1440)$&&$N(^2S^\prime_S)_{\frac 12^+}$&&1.440&&0.150\\[1ex]    
$P_{11}(1710)$&&$N(^2S_M)_{\frac 12^+}$&&1.710&&0.100\\[1ex]    
$P_{13}(1900)$&&$N(^2D_M)_{\frac 32^+}$&&1.900&&0.500\\[1ex]    
$F_{15}(2000)$&&$N(^2D_M)_{\frac 52^+}$&&2.000&&0.490\\[1ex]      
\end{tabular}
\end{center} 
\end{table}
%
%%%%%%%%%%%%%%%%%%%%%%%%%%%%%%%%%%%%%%%%%%%%%%%%%%%% 

One of the main reasons that the $SU(6)\otimes O(3)$ symmetry is
broken is due to the configuration mixings caused by the one gluon
exchange~\cite{IKK_78}. 
Here, the most relevant configuration mixings are those of the
two $S_{11}$ and the two $D_{13}$ states around 1.5 to 1.7 GeV. The 
configuration mixings can be expressed in terms of the mixing angle
between the two $SU(6)\otimes O(3)$ states $|N(^2P_M)>$  and 
$|N(^4P_M)>$, with the total quark spin 1/2 and 3/2;  
%
%%%
%
\begin{eqnarray}\label{eq:MixS}
\left(\matrix{|S_{11}(1535)> \cr
|S_{11}(1650)>\cr}\right) &=&
\left(\matrix{ \cos \theta _{S} & -\sin \theta _{S}\cr
\sin \theta _{S} & \cos \theta _{S}\cr}\right) \nn \\
&& \left(\matrix{|N(^2P_M)_{{\frac 12}^-}> \cr
|N(^4P_M)_{{\frac 12}^-}>\cr}\right),  
\end{eqnarray}
%
%%%
%
and  
%
%%%
%
\begin{eqnarray}\label{eq:MixD}
\left(\matrix{|D_{13}(1520)> \cr
|D_{13}(1700)>\cr}\right) &=&
\left(\matrix{ \cos \theta _{D} & -\sin \theta _{D}\cr
\sin \theta _{D} & \cos \theta _{D}\cr}\right) \nn \\
&& \left(\matrix{|N(^2P_M)_{{\frac 32}^-}> \cr
|N(^4P_M)_{{\frac 32}^-}>\cr}\right),  
\end{eqnarray}
%
%%%
%
where the mixing angle $\theta$ is predicted to be $-32^\circ$
for the $S_{11}$ resonances and $6^\circ$ for the $D_{13}$ resonances
in the Isgur-Karl Model~\cite{IK_77}. 

To show how the coefficients $C_{N^*}$ are related to the mixing angles, 
we express the amplitudes ${\mathcal A}_{N^*}$ in terms of the 
product of the photo and meson transition amplitudes
\begin{eqnarray}\label{eq:MixAR}
{\mathcal A}_{N^*} \propto <N|H_m| N^*><N^*|H_e|N>,
\end{eqnarray}
where $H_m$ and $H_e$ are the meson and photon transition operators,
respectively. Using Eqs.~(\ref{eq:MixS}) to~(\ref{eq:MixAR}), 
for the resonance ${S_{11}(1535)}$ one finds
\begin{eqnarray}\label{eq:MixAS1}
{\mathcal A}_{S_{11}} &\propto& 
<N|H_m (\cos \theta _{S}
 |N(^2P_M)_{{\frac 12}^-}> - 
\sin \theta _{S} 
\nonumber\\
&&|N(^4P_M)_{{\frac 12}^-}>) 
 (\cos \theta _{S} <N(^2P_M)_{{\frac 12}^-}| -
\nonumber\\
&&
\sin \theta _{S} <N(^4P_M)_{{\frac 12}^-}|)  
 H_e|N>,
\end{eqnarray}
%%%%%%%%%%%%%%%%%%%%%%%%%%%%%%

Due to the Moorhouse selection rule~\cite{Moor},
the photon transition amplitude $<N(^4P_M)_{{\frac 12}^-}|H_e|N>$
vanishes~\cite{zpl97} in our model.
So, Eq.~(\ref{eq:MixAS1}) becomes
\begin{eqnarray}\label{eq:MixAS2}
{\mathcal A}_{S_{11}}&\propto& (\cos^2 \theta _{S} - {\cal {R}}
\sin 
\theta _{S}\cos \theta _{S}) 
<N|H_m|N(^2P_M)_{{\frac 12}^-}>
\nonumber\\
&& <N(^2P_M)_{{\frac 12}^-}|H_e|N>,
\end{eqnarray}
where $<N|H_m|N(^2P_M)_{{\frac 12}^-}> <N(^2P_M)_{{\frac 12}^-}|H_e|N>$
determines~\cite{zpl97} the CGLN amplitude for the 
$|N(^2P_M)_{{\frac 12}^-}> $ state, and the ratio
\begin{eqnarray}\label{eq:MixR}
{\cal {R}} =  \frac {<N|H_m|N(^4P_M)_{{\frac 12}^-}>}
{<N|H_m|N(^2P_M)_{{\frac 12}^-}>},
\end{eqnarray} 
is
a constant determined by the $SU(6)\otimes O(3)$ symmetry. Using the 
the meson transition operator $H_m$ from the Lagrangian used in deriving 
the CGLN amplitudes in the quark model, we find ${\cal {R}}$~=~-1 for the $S_{11}$
resonances and $\sqrt{1/10}$ for the $D_{13}$ resonances.
Then, the configuration mixing coefficient can be related to the
configuration mixing angles 
\begin{eqnarray}
C_{S_{11}(1535)} &=& \cos {\theta _{S}} ( \cos{\theta _{S}} - 
\sin{\theta _{S}}),\label{eq:MixS15} \\
C_{S_{11}(1650)} &=& -\sin {\theta _{S}} (\cos{\theta _{S}} + 
\sin{\theta _{S}}),\label{eq:MixS16} \\
C_{D_{13}(1520)} &=& \cos \theta _{D} (\cos\theta _{D} - 
\sqrt {1/10}
\sin\theta _{D}),\label{eq:MixD15} \\
C_{D_{13} (1700)} &=& \sin \theta _{D} (\sqrt {1/10}\cos\theta _{D} + 
 \sin\theta _{D}).\label{eq:MixD17}
\end{eqnarray}

\subsection{Photo-excitation helicity amplitudes and $\eta N$ decay width of 
baryon resonances}
\label{sec:Theory2}

The total cross section in the $\eta$ 
photoproduction for a given resonance can be expressed as 
\begin{eqnarray}\label{eq:Sig1}
\sigma \propto \Gamma_{\eta N} (A_{1/2}^2 + A_{3/2}^2).
\end{eqnarray}
In the quark model, the  
helicity amplitudes $(A_{1/2})_{qm}$ and  $(A_{3/2})_{qm}$
and the partial width $\Gamma_{\eta N}^{qm}$ 
are {\it calculated explicitly} (Tables~2 and~3).

Then the above configuration mixing 
coefficients $C_{N^*}$ are introduced and their numerical values are  
extracted by fitting the experimental data, so that
\begin{eqnarray}\label{eq:Sig2}
\sigma \propto \Gamma_{\eta N}^{th} (A_{1/2}^2 + A_{3/2}^2)_{qm},
\end{eqnarray}
where
\begin{eqnarray}\label{eq:Gam1}
\Gamma_{\eta N}^{th} \equiv C_{N^*}^2 \Gamma_{\eta N}^{qm}.
\end{eqnarray}
The purpose of the procedure developed here is to extract the 
experimental value of the partial width $\Gamma_{\eta N}^{exp}$ in
\begin{eqnarray}\label{eq:Sig3}
\sigma \propto\Gamma_{\eta N}^{exp} (A_{1/2}^2 + A_{3/2}^2)_{exp}.
\end{eqnarray}
Then from Eqs.~(\ref{eq:Sig2}) to (\ref{eq:Sig3}), 
\begin{eqnarray}\label{eq:Gam2}
\Gamma_{\eta N}^{exp}= C_{N^*}^2 \Gamma_{\eta N}^{qm} \frac {
 (A_{1/2}^2 + A_{3/2}^2)_{qm}}{(A_{1/2}^2 + A_{3/2}^2)_{exp}}.
\end{eqnarray}

As mentioned above,
the quantities $\Gamma^{qm}_{\eta N}$, ($A_{1/2}$)$_{qm}$ and 
($A_{3/2}$)$_{qm}$ 
in Eq.~(\ref{eq:Gam2}) can be explicitly calculated in the quark model, and 
consistency requires that the Lagrangian used in evaluating these 
quantities must be the same as that in deriving the CGLN amplitudes for 
each resonance~\cite{zpl97}. The resulting photon vertex from the Lagrangian used in 
deriving the CGLN amplitudes is slightly different from those used in the previous
calculations\cite{KI-80,LC90}. As we will show later, this 
does not lead to significant changes
in the numerical results.  The derivation of the helicity amplitudes is standard, 
and we give them
in Table~2 for the process ${N^*}\to \gamma p$.

We would like to underline that the present quark model approach within
the $SU(6)\otimes O(3)$ symmetry limit, predicts vanishing values for
the $\gamma p$ photo-decay amplitudes for the resonances
$D_{13}(1700)$ and $D_{15}(1675)$. 
In our previous work~\cite{LS-1}, in order to investigate possible 
deviations from this symmetry,
we used the same expressions for the $D_{13}(1700)$ resonance as for the 
$D_{13}(1520)$ due to the configuration mixing effects.
 In the case of the resonance $D_{15}(1675)$, the configuration mixing effect is
very small since there is only one $D_{15}$ configuration in this mass region. Thus, 
for this latter resonance, the helicity 
amplitudes presented in the Table 2 correspond to the CGLN amplitudes
for the $\gamma n \to \eta n$ channel, which was discussed in more detail 
in our previous study~\cite{LS-1}. In this work, we have adopted the same procedure.

Finally, the formula derived within our quark model approach for the 
resonance decaying into the $\eta N$ are summarized
in Table~3. Here also we have consistently used
the same Lagrangian as that in deriving the CGLN amplitudes 
in Ref.~\cite{zpl97}.

\onecolumn
%%%%%%%%%%%%%%%%%%%%%%%%%% TABLE 2
%
\begin{table}[htb]\label{tab:Hel}
\caption{Electromagnetic helicity amplitudes for the $\gamma p$ within the 
present quark model, with 
$E_{\gamma}$ the energy of the incoming photon,
$m_q$ = 330 MeV quark mass,
%
%$\mu$,
%
and $e^{-{\mathcal{K}^2}/{6\alpha_{ho}^2}}$ a form factor in the
harmonic 
oscillator basis.
Here
${\mathcal{K}} = (\frac { \sqrt {2s} M_N} {s+M_N^2}) k$,
with $M_N$ the mass of the nucleon. 
As explained in the text, for the $D_{15}$ the $\gamma n$ helicity 
amplitudes are given.
}
\begin{center}
\begin{tabular}{lll}
\hline
\multicolumn{1}{c}{Resonance}&\multicolumn{1}{c}{$A^p_{1/2}$}
&\multicolumn{1}{c}{$A^p_{3/2}$} \\ \hline
%    Resonance & $A^p_{1/2}$ &  $A^p_{3/2}$ \\ \hline
%
$S_{11}$ & 
$\frac {2\sqrt{2}}3 \left (\frac {E_{\gamma} m_q}{\alpha_{ho}^2} + 
\frac 12\frac {{\mathcal{K}}^2}{\alpha_{ho}^2}\right ) 
\sqrt {\frac {\pi}{E_{\gamma}}} \mu
\alpha_{ho} 
e^{-\frac {{\mathcal{K}}^2}{6\alpha_{ho}^2}}$  &    \\[1ex]
$P_{11}$ & 
$- \frac 1{3\sqrt{6}}\left (\frac {\mathcal{K}} {\alpha_{ho}} \right)^2
\sqrt {\frac {\pi}{E_{\gamma}}} \mu {\mathcal{K}} 
e^{-\frac {{\mathcal{K}}^2}{6\alpha_{ho}^2}}$  &    \\[1ex]
$P_{13}$ & 
$\frac {2}{\sqrt{15}} \left (\frac 
{E_{\gamma} m_q}{\alpha_{ho}^2} +\frac 13 
\frac {{\mathcal{K}}^2}{\alpha_{ho}^2}\right ) 
\sqrt {\frac {\pi}{E_{\gamma}}} \mu {\mathcal{K}} 
e^{-\frac {{\mathcal{K}}^2}{6\alpha_{ho}^2}}$  &
$-\frac 2{3\sqrt{5}} \frac {E_{\gamma} m_q}{\alpha_{ho}^2}  
\sqrt {\frac {\pi}{E_{\gamma}}} \mu {\mathcal{K}}
e^{-\frac {{\mathcal{K}}^2}{6\alpha_{ho}^2}}$    \\[1ex]
$D_{13}$ & 
$\frac 23 \left (\frac {E_{\gamma} m_q}{\alpha_{ho}^2} - 
\frac {{\mathcal{K}}^2}{\alpha_{ho}^2}\right ) 
\sqrt {\frac {\pi}{E_{\gamma}}} \mu \alpha_{ho} 
e^{-\frac {{\mathcal{K}}^2}{6\alpha_{ho}^2}} $  &
$\frac 2{\sqrt{3}} \frac {E_{\gamma} m_q}{\alpha_{ho}^2}  
\sqrt {\frac {\pi}{E_{\gamma}}} \mu \alpha_{ho} 
e^{-\frac {{\mathcal{K}}^2}{6\alpha_{ho}^2}} $    \\[1ex]
$D_{15}$ & 
$\frac {-2}{3\sqrt{10}} \frac {{\mathcal{K}}^2}{\alpha_{ho}^2}
\sqrt {\frac {\pi}{E_{\gamma}}} \mu \alpha_{ho} 
e^{-\frac {{\mathcal{K}}^2}{6\alpha_{ho}^2}} $  &
$\frac {-2}{3\sqrt{5}} \frac {{\mathcal{K}}^2}{\alpha_{ho}^2}
\sqrt {\frac {\pi}{E_{\gamma}}} \mu \alpha_{ho} 
e^{-\frac {{\mathcal{K}}^2}{6\alpha_{ho}^2}} $    \\[1ex]

$F_{15}$ &
$\frac {2\sqrt{2}}{3\sqrt{5}} \left (\frac 
{E_{\gamma} m_q}{\alpha_{ho}^2} -\frac 12 
\frac {{\mathcal{K}}^2}{\alpha_{ho}^2}\right ) 
\sqrt {\frac {\pi}{E_{\gamma}}} \mu {\mathcal{K}} 
e^{-\frac {{\mathcal{K}}^2}{6\alpha_{ho}^2}}$   &
$\frac 4{3\sqrt{5}} \frac {E_{\gamma} m_q}{\alpha_{ho}^2}  
\sqrt {\frac {\pi}{E_{\gamma}}} \mu {\mathcal{K}}
e^{-\frac {{\mathcal{K}}^2}{6\alpha_{ho}^2}}$    \\[1ex]
\hline
\end{tabular}
\end{center} 
\end{table}
%%%%%%%%%
%
%
%\newpage
%

%\end{document}
%%%%%%%%%%%%%%%%%%%%%%%%%% TABLE 3
%
\begin{table}[htb]\label{tab:Widt}
\caption{Expressions for the $\eta N$ decay widths of the resonances,
with ${\mathcal{Q}} = (\frac { M_N} {E_f}) q$,
and $E_f$  the energy of the final state nucleon.
}
\begin{center}
\begin{tabular}{ll}
\hline
\multicolumn{1}{c}{Resonance}&\multicolumn{1}{c}{$\Gamma_{\eta N}^{qm}$} \\ 
\hline
$S_{11}$ & 
$\frac 4{9} \alpha_{\eta NN} 
\frac {E_f}{M_{N^*}} \frac {\mathcal{Q}}{M_N^2} 
 \left [ \frac {M_{N^*}+M_{N}}{E_f+M_N} 
\frac {{\mathcal{Q}}^2}{\alpha_{ho}}- \frac 3 2 \frac {E_\eta}
{m_q} \right ]^2 
 e^{-\frac {{\mathcal{Q}}^2}{3\alpha_{ho}^2}}$ 
\\[1ex]
$P_{11}$ & 
$\frac 2{3 }
 \alpha_{\eta NN} \frac {E_f}{M_{N^*}} \frac {{\mathcal{Q}}}{M_N^2}
 \left [ \frac {M_{N^*}+M_{N}}{E_f+M_N} 
\frac {{\mathcal{Q}}^3}{\alpha_{ho}^2}- \frac {E_\eta {\mathcal{Q}}}
{m_q} \right ]^2 
 e^{-\frac {{\mathcal{Q}}^2}{3\alpha_{ho}^2}}$ 
\\[1ex]
$P_{13}$ & 
$\frac 1{15} \alpha_{\eta NN} 
\frac {E_f}{M_{N^*}} \frac {\mathcal{Q}}{M_N^2} 
 \left [ \frac {M_{N^*}+M_{N}}{E_f+M_N} 
\frac {{\mathcal{Q}}^3}{\alpha_{ho}^2}-\frac 5 2 \frac {E_\eta {\mathcal{Q}}}
{m_q} \right ]^2 
 e^{-\frac {{\mathcal{Q}}^2}{3\alpha_{ho}^2}}$
\\[1ex]
$D_{13}$ & 
$\frac 4{9} \alpha_{\eta NN} 
\frac {E_f}{M_{N^*}} \frac {\mathcal{Q}}{M_N^2} 
 \left [ \frac {M_{N^*}+M_{N}}{E_f+M_N} 
\frac {{\mathcal{Q}}^2}{\alpha_{ho}}\right ]^2 
 e^{-\frac {{\mathcal{Q}}^2}{3\alpha_{ho}^2}}$ 
\\[1ex]
$D_{15}$ & 
$\frac 4{15} \alpha_{\eta NN} 
\frac {E_f}{M_{N^*}} \frac {\mathcal{Q}}{M_N^2} 
 \left [ \frac {M_{N^*}+M_{N}}{E_f+M_N} 
\frac {{\mathcal{Q}}^2}{\alpha_{ho}}\right ]^2 
 e^{-\frac {{\mathcal{Q}}^2}{3\alpha_{ho}^2}}$ 
\\[1ex]

$F_{15}$ & 
$\frac 1{15}
 \alpha_{\eta NN} \frac
{E_f}{M_{N^*}} \frac {{\mathcal{Q}}}{M_N^2}
\left [ \frac {M_{N^*}+M_{N}}{E_f+M_N}
\frac {{\mathcal{Q}}^3}{\alpha_{ho}^2}\right ]^2 
  e^{-\frac {{\mathcal{Q}}^2}{3\alpha_{ho}^2}}$
\\[1ex]
\hline
\end{tabular}
\end{center} 
\end{table}
\twocolumn
%%%%%%%%%%%%%%%%%%%%%%%%%%%%%%%%%%%%%%%%%%%%%%%%%%%%%%%%%%%%%%%%%%%%%
%
%%%%%%%%%%%% Section III
%
%-------------------- RESULTS -----------------------------------
% 
\section{Results and Discussion}
\label{sec:Results}
In this Section, we compare the results of the quark model
presented above, with the recent 
data~\cite{Mainz,elsa,graal,Graal2000,Fred}.
%
%%%%%%%%%%%% 
%
\subsection{Fitting procedure and extracted parameters}
\label{sec:Results1}
As mentioned above, within the exact $SU(6)\otimes O(3)$ symmetry
scheme, the only free parameters of our approach are: the strength
of the harmonic oscillator $\alpha_{ho}$ and the $\eta NN$ coupling 
constant $\alpha_{\eta NN} \equiv 2 g_{\eta NN}$.
However, introducing the symmetry breaking effects {\it via}
the $C_{N^*}$ coefficients (Eq.~\ref{MRexp}), we need in addition
one free parameter per resonance. Given recent results from 
Graal~\cite{Graal2000} and JLab~\cite{Armstrong,CLAS_e},
we leave also as free parameters the mass and the width of the
dominant $S_{11}(1535)$ resonance.

In this Section, we report on three models summarized in Table~4
and described below:
\vskip 3 truemm

a) {\bf Model A:} This model includes all the eleven known relevant 
resonances (Table~1)
with mass below 2 GeV, and hence contains 14 free parameters. Note that
the strength of the Roper resonance is kept at its quark model value
$C_{P11(1440)}$ = 1., as discussed in Ref.~\cite{LS-1}.
\vskip 2 truemm

b) {\bf Model B:} Here we introduce the mixing angle constraints, 
Eqs.~(\ref{eq:MixS15}) to~(\ref{eq:MixD17}),
as explained in Section~(\ref{sec:Theory1}) So, the two strengths for 
the $S_{11}(1535)$
and $S_{11}(1650)$ resonances are replaced by the mixing angle $\theta_S$.
This is also the case for the resonances $D_{13}(1520)$ and
$D_{13}(1700)$ related by the  mixing angle $\theta_D$. The number of 
free parameters is then reduced to 12.
\vskip 2 truemm

c) {\bf Model C:} In the presence of the mixing angle constraints as above,
and for the reasons that will be explained in 
Section~(\ref{sec:Results2}), we introduce
a third S11 resonance with three free parameters; namely, its mass, width, and
strength. The number of free parameters increases to 15.
\vskip 3 truemm

The free parameters of all the above three models have been extracted
(Table~4) using the MINUIT minimization code~\cite{MINUIT} from the 
CERN Library.
The fitted data base contains 400 values: differential 
cross-sections from Mainz~\cite{Mainz} and Graal~\cite{Graal2000}, 
and the beam asymmetry polarization data from Graal~\cite{graal}.

In the following, we compare the results of our models with different 
fitted observables,
but also with predicted ones, namely, total cross section and 
the polarized target asymmetry.

\onecolumn
%
%%%%%%%%%%%%%%%%%%%%%%%%%%%%  TABLE IV
%
\begin{table}\label{tab:Models}
\caption{Free parameters and their extracted values; masses and widths are given in GeV.}
\begin{center}
\begin{tabular}{lccc}  
\hline
&  &  & \\
Parameter& Model A & Model B & Model C \\
&  &  & \\
\hline
&  &  & \\
 $\alpha_{ho}^2$ (GeV$^2)$   & 0.090 $\pm$ 0.001 & 0.090 $\pm$ 0.001 & 0.090 $\pm$ 0.007  \\
$\alpha_{\eta NN} \equiv 2 g_{\eta NN}$& 0.898 $\pm$ 0.012 & 1.530 $\pm$ 0.015 & 1.467 $\pm$ 0.020  \\[8pt]
%&  &  & \\
 Mass of $S_{11}(1535)$ & 1.530 $\pm$ 0.001 & 1.530 $\pm$ 0.001 & 1.542 $\pm$ 0.001  \\
 Width of $S_{11}(1535)$ & 0.140 $\pm$ 0.001 & 0.142 $\pm$ 0.001 & 0.162 $\pm$ 0.001  \\[8pt]
% &  &  & \\
$C_{S_{11}(1535)}$  & 1.500 $\pm$ 0.001 & {\it 1.167 $\pm$ 0.009} & {\it 1.120 $\pm$ 0.003}  \\
 $C_{S_{11}(1650)}$ & -0.182 $\pm$ 0.011 & {\it-0.16 7$\pm$ 0.009}  & {\it -0.200 $\pm$ 0.003}  \\
 $\theta_S$      &          -         &-32.2$^\circ$ $\pm$1.8$^\circ$& -26.6$^\circ$ $\pm$ 0.8$^\circ$ \\[8pt]
%
% &  &  & \\
 Mass of the third $S_{11}$ & - & - & 1.729 $\pm$ 0.003  \\
 Width of the third $S_{11}$& - &  - & 0.183 $\pm$ 0.010  \\
 Strength of the third $S_{11}$& - &-& 0.542 $\pm$ 0.053 \\[8pt]
%
% &  &  & \\
 $C_{D_{13}(1520)}$ & 1.500 $\pm$ 0.014 & {\it 0.964 $\pm$ 0.002} & {\it 0.964 $\pm$ 0.002}  \\
$C_{D_{13}(1700)}$ & 0.100 $\pm$ 0.005 & {\it 0.036 $\pm$ 0.002} & {\it 0.036 $\pm$ 0.002}  \\
 $\theta_D$ &          -        & 5.1$^\circ$ $\pm$0.2$^\circ$ & 5.1$^\circ$ $\pm$ 0.2$^\circ$ \\[8pt]
% &  &  & \\
% P11(1440)           & 1.0            & 1.0 & 1.0  \\
 $C_{P_{11}(1710)}$  & 1.790 $\pm$ 0.385 &-0.837 $\pm$ 0.449 &-1.057 $\pm$ 0.206  \\[8pt]
% &  &  & \\
 $C_{P_{13}(1720)}$ & 0.053 $\pm$ 0.052&0.305 $\pm$ 0.058& 1.000 $\pm$ 0.010  \\
 $C_{P_{13}(1900)}$ &-2.500 $\pm$ 0.030 &-2.500 $\pm$ 0.013 & -2.478 $\pm$ 0.047  \\[8pt]
% &  &  & \\
 $C_{F_{15}(1688)}$ & 0.814 $\pm$ 0.241 & 0.761 $\pm$ 0.202 & 2.123 $\pm$ 0.102  \\
 $C_{F_{15}(2000)}$ &-2.500 $\pm$ 0.028 &-2.500 $\pm$ 0.026 & 0.201 $\pm$ 0.426  \\[8pt]
% &  &  & \\
 $C_{D_{15}(1675}$ &-0.505 $\pm$ 0.030 & -0.382 $\pm$ 0.022 & -0.169 $\pm$ 0.024  \\[8pt]
\hline
 &  &  & \\
 $\chi^2_{d.o.f.}$ & 3.2 & 3.8 & 1.6 \\
 &  &  & \\
\hline
\end{tabular}
\end{center}
\end{table}
%
%%%%%%%%%%%%%%%%%%%%%%%%%% 
%

\twocolumn

\subsection{Differential Cross-Section}
\label{sec:Results2}
The recent and accurate data for the differential cross-sections come from two groups
and have been included in the fitted data base.

i) Mainz data~\cite{Mainz}: Angular distributions, $\theta^{cm}_{\eta}~\approx$ 
$26^\circ$ to $154^\circ$, have been reported between $E^{lab}_{\gamma}$ =
0.716 GeV and 0.790 GeV at 10 energies. This data base contains 100 data points.

ii) Graal data~\cite{Graal2000}: Here, the angular distributions, 
$\theta^{cm}_{\eta}~\approx$ 
$20^\circ$ to $160^\circ$, have been measured between $E^{lab}_{\gamma}$ =
0.714 GeV and 1.1 GeV at 24 energies. This data base contains 225 data points.
This set of data has larger uncertainties than the Mainz data, but goes well
above the first resonance region.
%%%%%%%%%%%%%%%%%%%%%%
%%%%%%%%%%%%%%%%%%%%%%  FIG 1 %%%%%%%%%%%%%%%%%%%%%%
%%%%%%%%%%%%%%%%%%%%%%
\begin{figure*}
\begin{center}
%\rule{5cm}{0.2mm}\hfill\rule{5cm}{0.2mm}
%\vskip 2.5cm
%\rule{5cm}{0.2mm}\hfill\rule{5cm}{0.2mm}
\psfig{figure=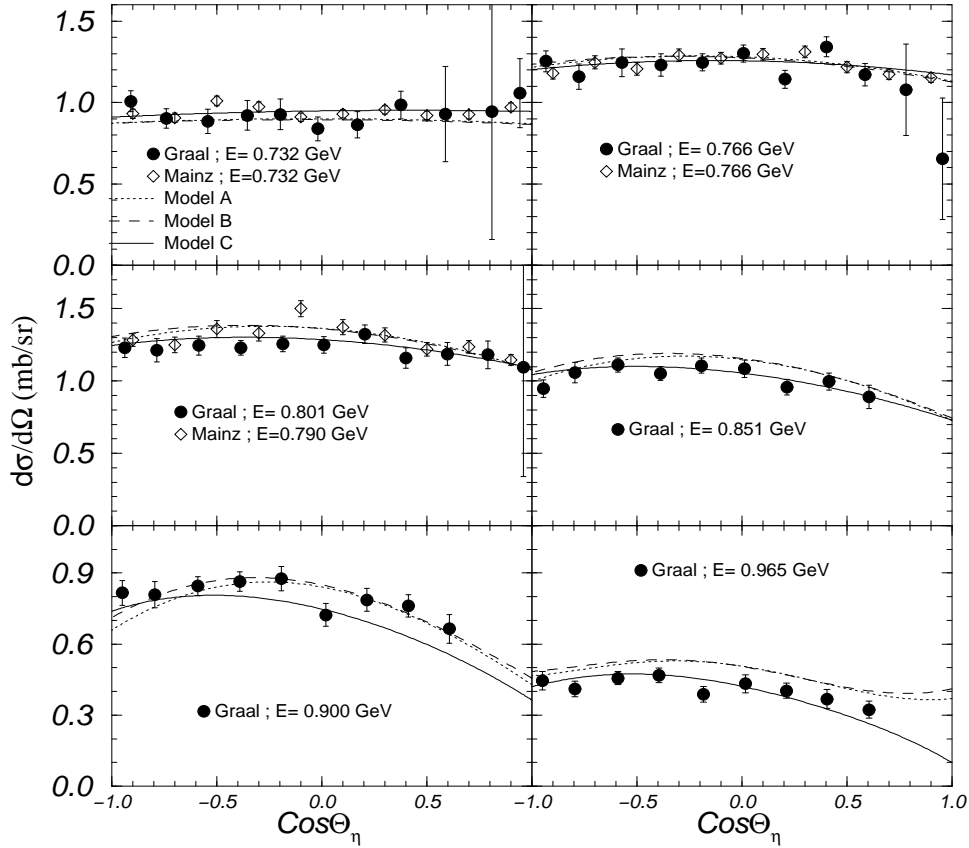,width=15.cm,height=12.cm}
\caption{Differential cross section for the process 
$\gamma p \to \eta p$: angular distribution for 
$E_{\gamma}^{\rm lab}$ = 0.732 GeV to 0.965 GeV.
The curves come from the models A (dotted), B (dashed),
and C (full).
Data are from Refs.~\cite{Mainz}  (empty diamonds), and~\cite{Graal2000} (full circles).}
\label{fig:dsigma1}
\end{center}
\end{figure*}
%%%%%%%%%%%%%%%%%%%%%%%%%%%%%%%%%%

Between the two above data sets, there are four overlapping energies. Here, to
keep the number of the figures reasonable, we show comparisons at twelve energies.
In Figure~1, data and our results are shown between 0.732 GeV and 0.965 GeV, where
there are three overlapping energies between Mainz and Graal data. 
At the two lowest energies, the three models A, B, and C reproduce equally well
these data. At two intermediate energies as well as at the highest one, the model C 
turns out to be superior to the the models A and B.
%%%%%%%%%%%%%%%%%%%%%%
%%%%%%%%%%%%%%%%%%%%%%  FIG 2 %%%%%%%%%%%%%%%%%%%%%%
%%%%%%%%%%%%%%%%%%%%%%
\begin{figure*}
\begin{center}
%\rule{5cm}{0.2mm}\hfill\rule{5cm}{0.2mm}
%\vskip 2.5cm
%\rule{5cm}{0.2mm}\hfill\rule{5cm}{0.2mm}
\psfig{figure=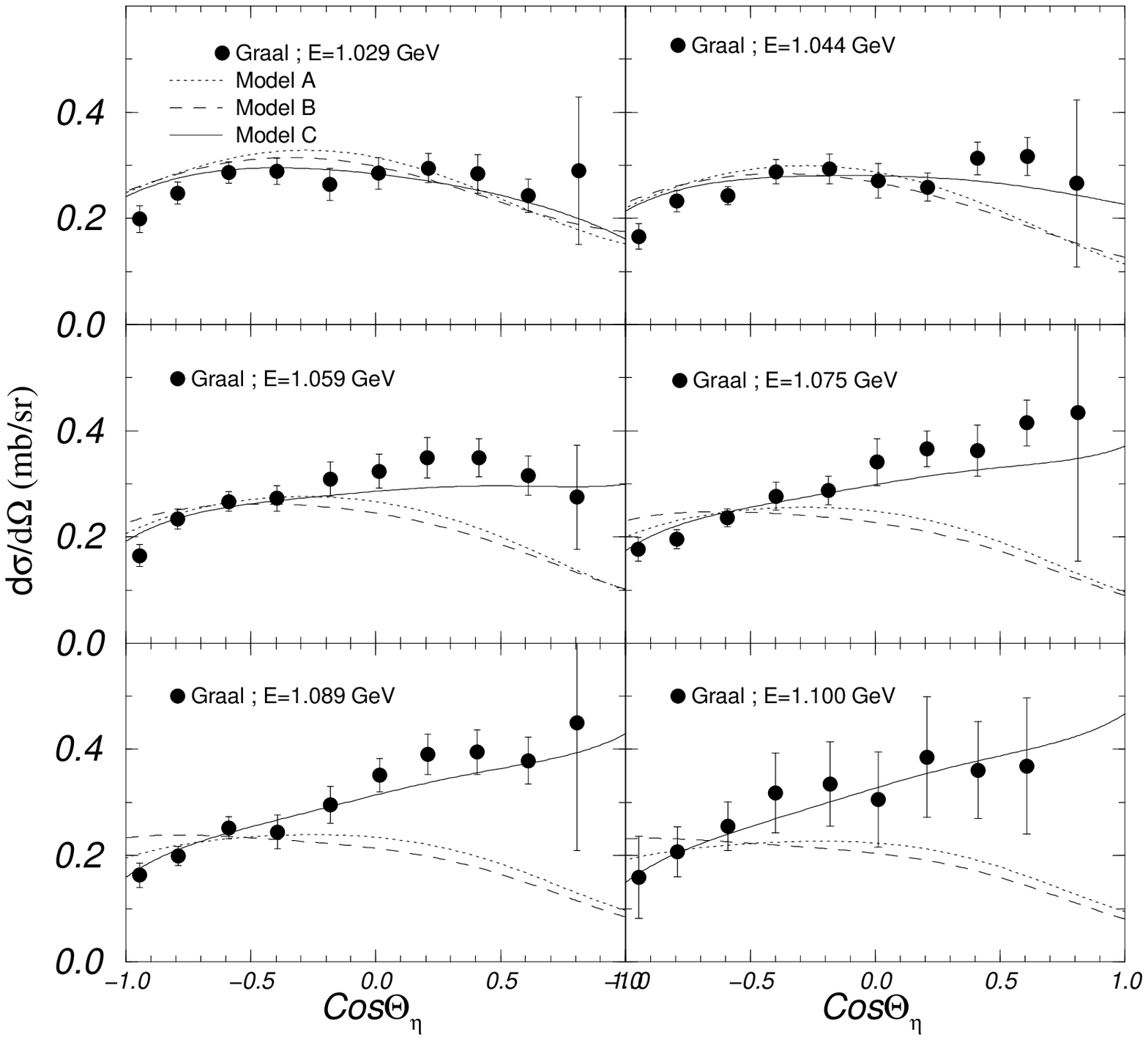,width=15.cm,height=12.cm}
\caption{Same as Fig.1, but for
$E_{\gamma}^{\rm lab}$ = 1.029 GeV to 1.1 GeV.}
\label{fig:dsigma2}
\end{center}
\end{figure*}
%%%%%%%%%%%%%%%%%%%%%%%%%%%%%%%%%%
%
In Figure~2, results from 1.0 to 1.1 GeV are depicted. With increasing energy, the
forward angle data are reproduced correctly only by the model~C, while above 
1.06 GeV they are badly reproduced by the models A and B. 
Before proceeding to comparisons with other observables, we discuss these three 
models in more detail.
\vskip 3 truemm

{\bf Model~A:}~Here the strengths of all relevant resonances are left as free parameters.
This is the most simple minded procedure in the sense that the $SU(6)\otimes O(3)$ 
symmetry breaking is introduced, allowing contributions from the three resonances
$S_{11}(1650)$, $D_{13}(1700)$, and $D_{15}(1675)$, without any constraint from
the mixing angle relations (Eqs.~(\ref{eq:MixS15}) to~(\ref{eq:MixD17})).

This approach was already applied in a previous paper~\cite{LS-1}, to a more restricted data base.
One of the main interests here is to find out by how much the strengths of those
three resonances deviate from zero, which is the predicted quark model value within
the exact $SU(6)\otimes O(3)$ symmetry. Table~4 shows that these deviations stay indeed
small for the $S_{11}(1650)$ and $D_{13}(1700)$. Compared to our previous 
work (see model M-7 in Table~2 of Ref.~\cite{LS-1}),
these coefficients decrease significantly due to a more copious data base.
The rather large extracted strength value for the $D_{15}(1675)$ will be discussed later.

The new data set from Graal~\cite{Graal2000} included in the present work, brings in another change
compared to our earlier work~\cite{LS-1}: the width of the $S_{11}(1535)$ goes down from 230 MeV to
about 140 MeV. This latter value, dictated by the higher energy part of the Graal data,
is compatible with the recent extractions from data~\cite{Graal2000,Armstrong}.
The strong correlations among this quantity, the harmonic oscillator strength,
the $\eta NN$ coupling, and the $S_{11}(1535)$ strength, explain the differences
between the extracted values in Ref.~\cite{LS-1} and the present model~A.

Finally, the mass of the $S_{11}(1535)$  comes out slightly smaller than its PDG value.
\vskip 2 truemm

{\bf Model~B:} A correct treatment of the $SU(6)\otimes O(3)$ symmetry breaking requires the
introduction of the mixing angles. We have used Eqs.~(\ref{eq:MixS15}) 
to~(\ref{eq:MixD17}) to replace, as free parameters,
the strengths of the $S_{11}(1535)$ and $S_{11}(1650)$ by $\theta_S$ and those
of the $D_{13}(1520)$ and $D_{13}(1700)$ by $\theta_D$. In Table~4, we give in {\it italic}
the values of those strengths using the extracted mixing angles and 
Eqs.~(\ref{eq:MixS15}) to~(\ref{eq:MixD17}). The absolute
values of all four strengths decrease compared to those of model~A. This is also the
case for the resonance $D_{15}(1675)$. The other significant changes concern
the strengths of the $P_{13}(1720)$ and $P_{11}(1710)$. This latter resonance plays however,
a minor role and hence its extracted strength bears large uncertainty. Note that the 
reduced $\chi^2$ increases by about 15\% compared to the model A, because of the additional
constraint on the configuration mixings.

Our extracted mixing angles are in agreement with the quark model predictions~\cite{IK_77}
and results coming from the large-$N_c$ effective field theory based 
approaches~\cite{Carlson,Pir98}. However, the model B
does not offer satisfactory features when compared to the data between threshold and
1.1 GeV (Figs.~1 and 2).
\vskip 2 truemm
{\bf Model C:} Results of the models A and B show clearly that an approach containing a
correct treatment of the Born terms and including {\it all known resonances} in the 
{\it s-} and {\it u-}channels {\it does not} lead to an acceptable model, even within
broken $SU(6)\otimes O(3)$ symmetry scheme. The forward peaking behavior in the 
differential cross section around 1.1 GeV suggests the presence of a large 
$S$-wave component that can not be accommodated by the known $S_{11}$ resonances.

To go further, one possible scenario is to investigate manifestations of yet undiscovered
resonances, because of their weak or null coupling to the $\pi N$ channel.
A rather large number of such resonances has been predicted by various authors.
To find out which ones could be considered as relevant candidates, we performed
a detail study of all the observables for which data are available and studied
their multipole structures. This investigation, the results of which will be
reported elsewhere, led us to the conclusion that a predicted~\cite{zr96} 
third $S_{11}$ resonance, with M=1.712 GeV and $\Gamma_{T}$=184 MeV, could
be an appropriate candidate. If there is indeed an additional $S$-wave resonance in this
mass region, its dependence on incoming photon and outgoing meson momenta would be 
qualitatively similar to that of the $S_{11}(1535)$, even though the form factor
might be very different. Thus, for this new resonance, we use the same CGLN
amplitude expressions as for the $S_{11}(1535)$. 
We left however, its mass and width, as well as
its strength, as free parameters. The extracted values are given in Table~4 and
show amazingly close values to those predicted by the authors of Ref.~\cite{zr96}.
Moreover, for the one star $S_{11}(2090)$ resonance~\cite{PDG}, the Zagreb group
coupled channel analysis~\cite{Sva1} produces the following values
M = 1.792 $\pm$ 0.023 GeV and $\Gamma_T$ = 360 $\pm$ 49 MeV.

The differential cross-sections are well reproduced (Figs. 1 and 2) with this model.
The reduced $\chi^2$ is greatly improved and goes down to 1.6.

The strength of the harmonic oscillator $\alpha_{ho}^2$ comes out the same
for the three models and agrees with the findings of Ref.~\cite{LC90}.

Introducing this third resonance, hereafter referred to as $S_{11}(1730)$, modifies
the extracted values for the parameters of the two other $S_{11}$ resonances.
The mass and width of the first $S_{11}$ resonance come out compatible with
their recent determination by the CLAS collaboration~\cite{Armstrong},
as well as with those of the Zagreb group coupled channel analysis~\cite{Sva2}.

Moreover, the strengths of the $P_{11}(1710)$, $P_{13}(1720)$, and $D_{15}(1675)$
get very close to their predicted values by the quark model based on the 
$SU(6)\otimes O(3)$ symmetry. The $F_{15}(1680)$ plays a non-negligible role and
the two highest mass resonances, $P_{13}(1900)$ and $F_{15}(2000)$, have marginal
contributions. 

Finally, our extracted value for the ${\eta NN}$ coupling constant 
$g_{\eta NN}/\sqrt{4\pi}$ = 0.2 is compatible with recent 
determinations~\cite{Tia94,Ben95} of this fundamental quantity. 
However, results from the RPI group~\cite{RPI95a} as well as
from the hadronic sector~\cite{hadrons} suggest
values roughly between 2 and 7. Some possible origins of such
discrepancies are discussed in the literature~\cite{discrep}.
%%%%%%%%%%%% 
%
\subsection{Total cross-section}
\label{sec:Results3}
%%%%%%%%%%%%%%%%%%%%%%
%%%%%%%%%%%%%%%%%%%%%%  FIG 3 %%%%%%%%%%%%%%%%%%%%%%
%%%%%%%%%%%%%%%%%%%%%%
\begin{figure*}
\begin{center}
%\rule{5cm}{0.2mm}\hfill\rule{5cm}{0.2mm}
%\vskip 2.5cm
%\rule{5cm}{0.2mm}\hfill\rule{5cm}{0.2mm}
\psfig{figure=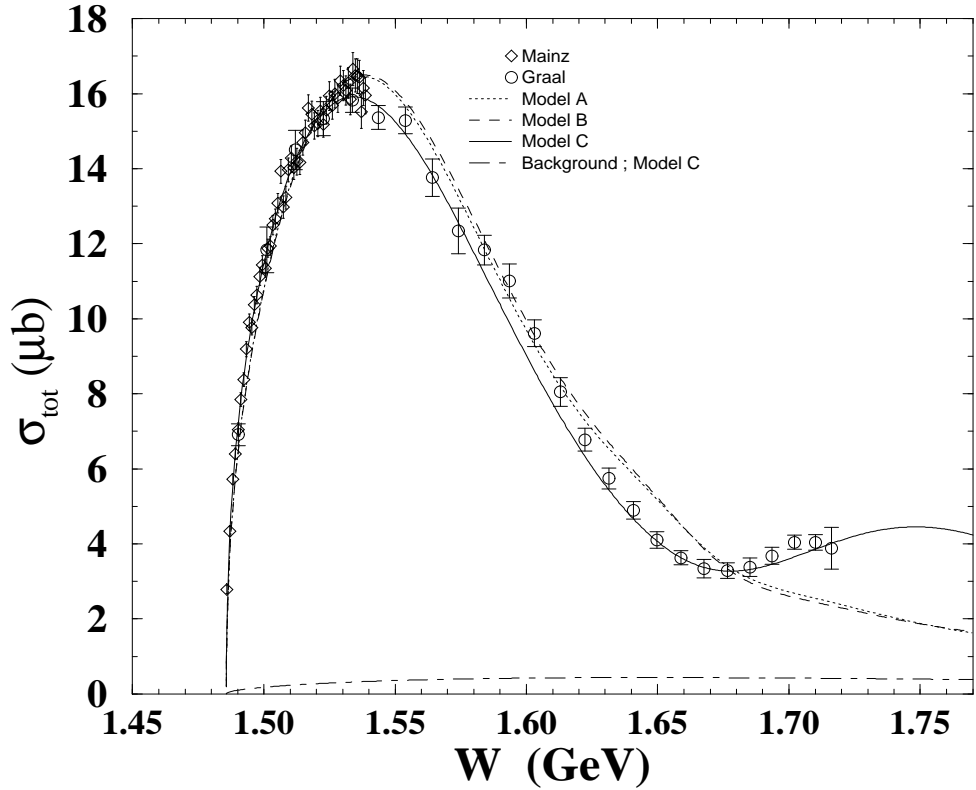,width=15.cm,height=12.cm}
\caption{Total cross section for the reaction 
$\gamma p \to \eta p$
as a function of total center-of-mass energy. The dot-dashed curve comes
from the background terms in model C, other curves and data are as in Fig.~1.}
\label{fig:sigtot}
\end{center}
\end{figure*}
%%%%%%%%%%%%%%%%%%%%%%%%%%%%%%%%%%
%
%
Figure~3 shows the results for the total cross-section. These data 
{\it were not}
included in the fitted data base. So, our curves can be considered as
semi-predictions. Here, the most striking feature is a minimum around
W=1.675 GeV ($E_{\gamma}^{lab} \approx$ 1.03 GeV), also reported by the
CLAS Collaboration~\cite{CLAS_e} in the $\eta$ electroproduction process.

Models A and B reproduce the data roughly up to W=1.62 GeV, missing badly the higher 
energy data. The introduction of the new resonance has a dramatic effect. The 
agreement between the curve C and the data is reasonable, and especially
the structure shown by data around W=1.7 GeV is nicely reproduced.
Note that, even the low energy data are better reproduced by model C than by
the two other models. Although the $S_{11}$(1730) resonance has a too high mass
to play a significant role close to threshold, its inclusion attributes to
the other two $S_{11}$ resonances more realistic roles. It is worthwhile
noting that the background terms contribution (Fig.~3) is small and
bears no structure.

Another striking feature is that the inclusion of the new resonance leads 
to higher extracted values for both mass and width of the $S_{11}(1535)$
resonance, compatible with the partial wave analysis~\cite{PDG,MS-92}
and a recent coupled channel~\cite{Sva2} results.
%%%%%%%%%%%% 
%
\subsection{Polarization observables}
\label{sec:Results4}
There are two sets of data for single polarization observables and we 
have investigated both.
\vskip 2 truemm

{\bf Polarized beam asymmetry:} The data come from the Graal collaboration~\cite{graal}
and contain 56 data points between 0.745 GeV and 1.09 GeV. 
In a previous work~\cite{LS-1}, we have performed a detailed
study of these data published in 1998. Results of a
more refined data analysis have been reported since then~\cite{Fred}.
In the present work, we have hence included these latter data in our fitted data base.
A challenging problem concerns the data at 1.057 GeV: the two most forward angle data,
at $\theta$ = 39$^\circ$ and to a less extent at 43$^\circ$, show an unexpected
increase. 
In Fig.~4, we show comparisons with data at this energy as well as at the two
adjacent ones, where the forward angle data are better reproduced. 
Although we do not settle the problem raised by those two
forward angle data points, we obtain a good description of the data, especially
with model~C.

For this observable, the quality of agreement with data at lower energies is 
comparable to that shown in Fig.~4, and to limit the number of figures, 
we do not show them here. 
%
%
%
%%%%%%%%%%%%%%%%%%%%%%
%%%%%%%%%%%%%%%%%%%%%%  FIG 4 %%%%%%%%%%%%%%%%%%%%%%
%%%%%%%%%%%%%%%%%%%%%%
\begin{figure}
\begin{center}
%\rule{5cm}{0.2mm}\hfill\rule{5cm}{0.2mm}
%\vskip 2.5cm
%\rule{5cm}{0.2mm}\hfill\rule{5cm}{0.2mm}
\psfig{figure=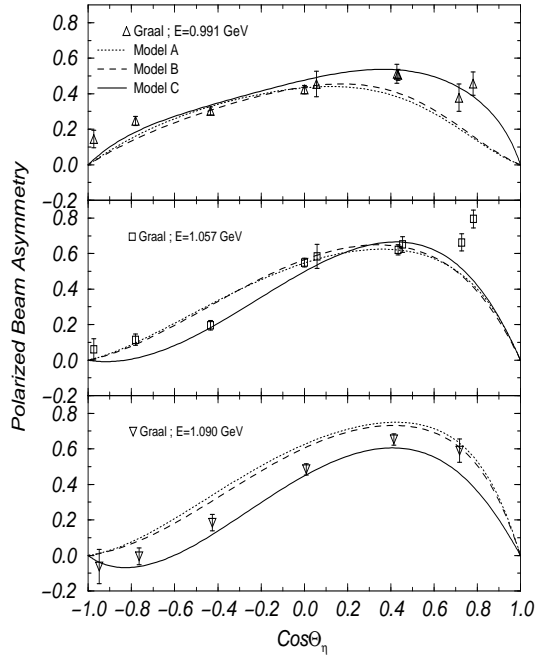,width=9.cm,height=9.cm}
\caption{Polarized beam asymmetry angular distributions for the reaction
${\gamma} p \to \eta p$. 
Curves are as in Fig.~1, and data from Refs.~\cite{graal,Graal2000,Fred}.}
\label{fig:Beam}
\end{center}
\end{figure}
%%%%%%%%%%%%%%%%%%%%%%%%%%%%%%%%%%
% 
%
\vskip 2 truemm

{\bf Polarized target asymmetry:} This observable has been measured at ELSA~\cite{elsa},
between 0.717 GeV and 1.1 GeV at 7 energies corresponding to 50 data points.

To evaluate the predictive power of our approach, we {\it did not} include these data
in our fitted data base. 
In Fig.~5, are depicted the results at six
measured energies with reasonable data points. In spite of the large 
experimental error bars, the superiority of the model~C in predicting this observable
is obvious. 

The nodal structure at low energies, seemingly indicated by the data, is however
not reproduced. This feature has already been discussed in detail in a previous
publication~\cite{LS-1}, and the conclusion presented there are not altered by
the models presented in this work. 

\onecolumn
%%%%%%%%%%%%%%%%%%%%%%
%%%%%%%%%%%%%%%%%%%%%%  FIG 5 %%%%%%%%%%%%%%%%%%%%%%
%%%%%%%%%%%%%%%%%%%%%%
\begin{figure}
\begin{center}
%\rule{5cm}{0.2mm}\hfill\rule{5cm}{0.2mm}
%\vskip 2.5cm
%\rule{5cm}{0.2mm}\hfill\rule{8cm}{0.2mm}
\psfig{figure=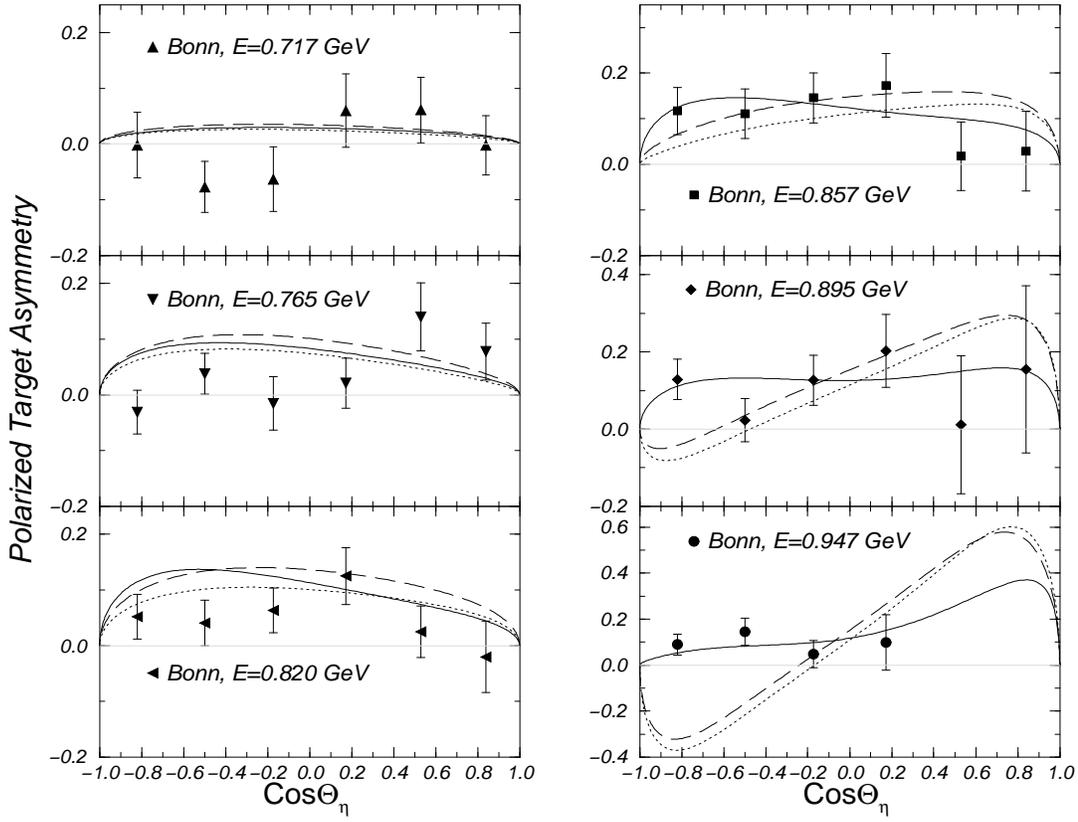,width=15.cm,height=12.cm}
\caption{Same as Fig.~4, but for polarized target asymmetry. 
Curves are as in Fig.~1, and data from Ref.~\cite{elsa}.}
\label{fig:Target}
\end{center}
\end{figure}
%%%%%%%%%%%%%%%%%%%%%%%%%%%%%%%%%%
%
\twocolumn
 
%%%%%%%%%%%% 
%
\subsection{Photo-excitation helicity amplitudes and partial decay widths}
\label{sec:Results5}
The process under investigation offers the possibility of
determining the electro-strong properties of the relevant
baryons. 
The connection between such properties and QCD-based (or inspired)
approaches has been emphasized by several 
authors~\cite{zr96,RPI95a,RPI97,zpl95}.

Up to now, given the state-of-the-art
for both theory and experiment, the investigations have been basically 
focused on the $S_{11}(1535)$ resonance.
In this Section we discuss first the case of this resonance, 
then we introduce the
relevant expressions for the $S_{11}(1650)$ resonance, before proceeding
to other ones in the first and second resonance region.

For the $S_{11}(1535)$ resonance, the quantities
of interest are the total width ($\Gamma_{T}$) of this 
latter resonance, and the electromagnetic helicity amplitude $A^p_{1/2}$.
Moreover, $A^p_{1/2}$ can be related
to the quantity $\xi$~\cite{RPI95b}, characteristic of the photo-excitation of 
the $S_{11}(1535)$ resonance and its decay into the  
$\eta$-nucleon channel, by the following relation:
\begin{equation}\label{eq:A}
A^p_{1/2}=
\sqrt{
\frac {q}{k} \frac {M_{R}}{M_p} 
\frac{\Gamma_{T}}{b_{\eta}}
}~ \xi
\end{equation}
and
\begin{eqnarray}\label{eq:xi}
\xi&=&\sqrt{\frac {\pi \alpha_{\eta} \alpha_{e} (E+M_{p}) }
{M_R^3}}
\frac {C_{S_{11}(1535)} \omega_{\gamma }}
{6 \Gamma_{T}}
\nonumber \\[2ex]
&&
\Bigl[\frac {2 \omega_{\eta}}{m_q}- \frac {2 q^2} {3 \alpha ^2}
( \frac {\omega_{\eta}} {E+M_p} +1) \Bigr]
(1+ \frac {k}{2m_q})
e^{-\frac 
{{k}^2+{q}^2}
{6\alpha^2}}.
\end{eqnarray}
For the branching ratio $b_\eta \equiv \Gamma_{\eta N} /\Gamma_{T}$, we
use 0.55~\cite{PDG,Armstrong}.

Given that the quark model used here predicts, in the $SU(6)\otimes O(3)$ 
symmetry limit, no contribution from the $S_{11}(1650)$ resonance, we 
cannot use the same approach as above for this latter resonance. 
However, we can derive the relevant expression for the partial width
of this resonance following Eqs.~(21) and (22) in Ref.~\cite{LS-1}.
This leads to the following relation where the Lorentz boost factor
(${\mathcal{K}}$ in Table~2) has been explicitly 
incorporated:
\begin{eqnarray}\label{eq:GamS}
\Gamma^{exp}_{{S_{11}(1650)} \to \eta N} &= &
\pi \alpha 
\Bigl[\alpha_{\eta NN} C^2_{S_{11}(1650)} \Bigr]
 \nn\\[2ex]
&&
\Bigl[\frac{1}{(A^p_{1/2})^2} \Bigr]^2 
\Bigl[\frac {2}{9}\frac {q}{k} 
\frac {M_N^3 {E^2_{\gamma }}}{ M_{N^*}^2 E^2_f} 
\frac {s (E_f+M_{N}) }{(s+M_N^2)^2} 
\Bigr]
 \nn\\[2ex]
&&
\Bigl[\frac {E_{\eta}}{m_q}- \frac {q^2} {3 \alpha^2_{ho}}
( \frac {E_{\eta}} {E_f+M_N} +1) \Bigr]^2
 \nn\\[2ex]
&&
(1+ \frac {k}{2m_q})^2
e^{-\frac{{k}^2+{q}^2}{3\alpha^2_{ho}}},
\end{eqnarray}
with $E_{\eta}$ the total energy of the outgoing $\eta$ meson.

For other resonances, we follow the expressions given in Tables~2 and 3.

For the resonances considered in this paper, the quark model results for 
electromagnetic helicity amplitudes and the 
latest PDG values~\cite{PDG} are listed in Tables 5 and 6. 

Our results for the first two $S_{11}$ resonances agree with the PDG values.

In the case of the $D_{13}(1520)$, 
both helicity amplitudes turn out compatible with those reported in the PDG.
This is also the case for the $A^p_{1/2}$ components of
the $D_{13}(1700)$, as well as for the
$A^p_{3/2}$ component of the $F_{15}(1680)$ resonances.

For the other amplitudes reported in Table 5, our results show significant 
deviations from the PDG values.
Such trends are also reported in the 
literature~\cite{Rev_CR-00,Cap92,FM,VPI,Per97,DRE-98}.

In the case of the $D_{15}(1675)$, as mentioned above, we extract the 
helicity amplitudes for the photon-neutron coupling. Then, we determine
those for the photon-proton coupling by using the following 
expressions~\cite{IKK_78}:
\begin{eqnarray}\label{eq:A3} 
A^p_{3/2} \approx -0.31 A^n_{3/2},
\end{eqnarray}
and
\begin{eqnarray}\label{eq:A1} 
A^p_{1/2} = (1/\sqrt {2}) A^p_{3/2}.
\end{eqnarray}

Our results are given in Table 6 and show good agreement with
the PDG values for all four amplitudes.
 
The extraction of the $\eta N$ decay width is straightforward: the 
coefficients $C_{N^*}$ for these resonances are given in Table 4  
and their masses and total decay widths in Table 1.
We present our numerical results for
the partial widths and branching ratios of the relevant resonances
in Table 7, where the second column gives the predictions 
of the quark model (see Table 3). 
The only uncertainty here comes from the coupling 
$\alpha_{\eta NN}$ $\equiv$ 2$g_{\eta NN}$ = 1.467$\pm$0.020.
The third column correspond to 
$\Gamma_{\eta N}^{th}=C^2_{N^*} \Gamma_{\eta N}^{qm}$, where another 
source of uncertainty is introduced because of the 
coefficients $C_{N^*}$ as reported in Table 4. In the fourth
column our values for the experimental width $\Gamma _{\eta N}^{exp}$,
as defined in Eq.~\ref{eq:Gam2}, are reported. For this quantity, the
major origin of the uncertainties comes from those in the helicity 
amplitudes as given in the PDG (see Table 5).
In the last column of Table 7, we give the branching ratio
BR = $\Gamma _{\eta N}^{exp} / \Gamma _{T}$, where the total widths
$\Gamma _{T}$ are taken from the PDG (see last column in 
Table 1). Our results for the $D_{13}(1520)$ are compatible 
with the width (0.6~MeV) reported in Ref.~\cite{Bijker}, but 
the branching ratio is larger than the values given
in Ref.~\cite{Tiator-99} (0.08$\pm$0.01 and 0.05$\pm$0.02).
For the $F_{15}(1680)$ resonance the only other available value comes,
to our knowledge, from an algebraic approach~\cite{BIL} which gives 
$\Gamma _{\eta N}$ = 0.5 MeV, much smaller than our result.

The uncertainties of the helicity amplitudes $[A_{1/2}]_{exp}$
and $[A_{3/2}]_{exp}$, 
extracted from experiments and reported in the PDG~\cite{PDG}, are major 
constraints on the determination of the  partial decay widths or
branching ratios within the present approach. 
For resonances with large experimental helicity amplitudes, 
such as the resonances $D_{13}(1520)$ and $F_{15}(1680)$,  
the uncertainties are small, so that the resulting errors 
in the $\eta N$ branching ratios are also small.  The extracted values for 
these two resonances are in good agreement with those in the PDG,  
showing the consistency of our approach.
However, for those resonances with smaller helicity amplitudes and 
larger uncertainties, such as the two P-wave resonances as well as the
$D_{13}(1700)$ and the $D_{15}(1675)$, the $\eta N$ decay width could not be 
well determined within our approach.  
Moreover, the rather small coefficients $C_{N^*}$ for these latter resonances 
obtained by fitting the photoproduction data in our previous study are 
due to the fact that their electromagnetic couplings are small, which is 
indeed consistent with the quark model predictions. 
However, our results here show that the corresponding $\eta N$ 
decay widths for these resonances {\it could be large}.
  
The above considerations show clearly the need for more comprehensive 
measurement of the $\eta$ photoproduction for {\it both} proton and neutron 
targets. The latter is especially desirable in investigation the
resonances $P_{11}(1710)$, $P_{13}(1700)$, $D_{13}(1700)$, and 
$D_{15}(1675)$, due to the fact that their electromagnetic couplings
 $\gamma n$ are larger than those for the proton target.
Therefore, their contributions to the $\eta$ photoproduction could be 
very significant.  

Finally, we would like to emphasize that the partial widths
extracted {\it via} a coupled channel T matrices analysis~\cite{Sva1} 
of the reactions $\pi N \to \eta N$ and $\eta N \to \eta N$ are
$\Gamma _{S_{11}(1650) \to \eta N}$ = 13 $\pm$ 7 MeV,
$\Gamma _{D_{13}(1520) \to \eta N}$ = 0.1 $\pm$ 0.1 MeV, and
$\Gamma _{F_{15}(1680) \to \eta N}$ = 0.2 $\pm$ 0.2 MeV. Within the
reported large uncertainties, the first two
values are compatible with our findings, while the width of 
the $F_{15}(1680)$ is significantly smaller than our result.

%\newpage
%
\onecolumn

%
%%%%%%%%%%%%%%%%%%%%%%%%%%%%%%% TABLE 4
%%
%%%%%%%%%%%%%%%%%%%%%%%%%%%%  TABLE V
%
%\begin{landscape} 
\begin{table}[ht]\label{tab:Hel1}
\caption{Photo-excitation helicity amplitudes in units of $10^{-3}$~GeV$^{-1/2}$.}
\begin{center}
\begin{tabular}{lccccccccccc}
\hline
\multicolumn{1}{c}{Resonance}
&\multicolumn{2}{c}{$A^{p}_{1/2}$}
&\multicolumn{1}{c}{{~~~~~~~}}
&\multicolumn{2}{c}{{$A^{p}_{3/2}$}}  \\
& {Model C}      & { PDG}  &   
& {Model C}      & { PDG}  \\
\hline 
{$S_{11}(1535)$}   & 64 & 90 $\pm$ 30 &&  &   \\ 
{$S_{11}(1650)$}   & 52 & 53 $\pm$ 16 &&  &   \\
{$P_{11}(1710)$}   &-36 & 9 $\pm$ 22 &&  &   \\  
{$P_{13}(1720)$}   & 156 & 18 $\pm$ 30  && -64 & -19 $\pm$ 20  \\
{$D_{13}(1520)$}   & -9 &-24 $\pm$ 9  && 149 & 166 $\pm$ 5  \\ 
{$D_{13}(1700)$}   &-21 &-18 $\pm$ 13 && 146 &  -2 $\pm$ 24  \\ 
{$F_{15}(1680)$}   & 34 &-15 $\pm$ 6  && 124 & 133 $\pm$ 12  \\
\hline     
\end{tabular}
\end{center} 
\end{table}
%\end{landscape}
%
%%%%%%%%%%%%%%%%%%%%%%%%%% TABLE VI
%
%
{\small {
\begin{table}[ht]\label{tab:Hel2}
\caption{Photo-excitation helicity amplitudes in units of $10^{-3}$~GeV$^{-1/2}$ for
the $D_{15}(1675)$ resonance.}
\begin{center}
\begin{tabular}{ccccccccccc}
\hline
\multicolumn{2}{c}{$A^{n}_{1/2}$}
&\multicolumn{2}{c}{{$A^{n}_{3/2}$}}
&\multicolumn{2}{c}{$A^{p}_{1/2}$}
&\multicolumn{2}{c}{{$A^{p}_{3/2}$}}  \\
 {Model C}      & { PDG}   
& {Model C}      & { PDG}  
& {Model C}      & { PDG}   
& {Model C}      & { PDG}\\
\hline  
&&&&&&\\  
-33 &-43 $\pm$ 12 & -46 & -58 $\pm$ 13&
10 & 19 $\pm$ 8 & 14 & 15 $\pm$ 9  \\
&&&&&\\
\hline   
\end{tabular}
\end{center} 
\end{table}
}}
%
%%%%%%%%%%%%%%%%%%%%%%%%%% TABLE 5
%%%%%%%%%%%%%%%%%%
% 
%%%%%%%%%%%%%%%%%%
%%
%%%%%%%%%%%%%%%%%%%%%%%%%%%%  TABLE VII
%  
\begin{table}[htb]\label{tab:Width}
\caption{$N^* \to N \eta$ decay widths (in MeV) and branching ratios from model C.}
\begin{center}
\begin{tabular}{cccccccc}
\hline
Resonance & $\Gamma _{\eta N}^{qm}$ & $\Gamma _{\eta N}^{th}$ & 
$\Gamma _{\eta N}^{exp}$ & BR$^{exp}$(\%)~~ \\
\hline 
$S_{11}(1650)$ &                 &                   &
4.0 $\pm$ 0.1 & 1.8 $\pm$  0.1 \\
$D_{13}(1520)$ &  ~0.7 $\pm$ 0.1 & ~0.6 $\pm$ 0.1~~~ & 0.5 $\pm$  0.2 & 
~0.4 $\pm$  0.1~~~ \\ 
$F_{15}(1680)$ &  ~6.5 $\pm$ 0.1 & 29.1 $\pm$ 1.3~~~ & 26.8 $\pm$ 9.4 & 
20.6 $\pm$  7.2~~~ \\
\hline     
\end{tabular}
\end{center} 
\end{table}
%
%%%%%%%%%%%%%%%

%
%
\twocolumn
%\newpage
%%%%%%%%%%%%%%%%%%%%%%%%%%
%-------------------- CONCLUSIONS -----------------------------------
\section{SUMMARY AND CONCLUSIONS}
\label{sec:Summary}
We reported here on a study of the process $\gamma p \to \eta p$ for 
$E_{\gamma}^{lab}$
between threshold and $\approx$ 1.2 GeV, using a chiral constituent quark approach. 

We extract the $\eta N$ branching ratio within our 
framework. The results for the $S_{11}(1650)$ and $D_{13}(1520)$ resonances
are compatible with the existing data. For the resonance $F_{15}(1680)$,  
as our earlier 
investigation~\cite{LS-1} showed, the strength of this resonance is very sensitive to the
polarization observables. Thus, more accurate data in this area are needed to 
confirm if this resonance has a large $\eta N$ branching ratio, as found in this
work.

We show how the symmetry breaking coefficients $C_{N^*}$ are expressed in terms of 
the configuration mixings in the quark model, thus establish a direct connection
between the photoproduction data and the internal quark gluon structure of baryon 
resonances. The extracted configuration mixing angles for the $S$ and $D$ wave resonances
in the second resonance region from a more complete data base are in good agreement 
with the Isgur-Karl model values~\cite{IK_77}, which predicted the configuration 
mixing angles based on the one gluon exchange~\cite{IKK_78}, as well as with results
coming from the large-$N_c$ effective field theory based 
approaches~\cite{Carlson,Pir98}.

However, one of the common features in our investigation of the $\eta$ photoproduction
at higher energies is that the existing S-wave resonance can not accommodate the
large S-wave component above $E_{\gamma}^{lab} \approx$ 1.0 GeV region. 
Thus, we introduce a third S-wave 
resonance in the second resonance region suggested in the literature~\cite{zr96}. 
The introduction of this new resonance, improves greatly
the quality of our fit and reproduces very well the cross-section increase
in the second resonance region. It even improves the agreement with low energy
data, by allowing the first region resonances to contribute in a more realistic
way. In particular, it describes very well the
forward peaking behavior compared to the models A and B, without the third S
wave resonance, which fail to generate the same trend.  
The quality of our semi-prediction for the total cross-section and our predictions
for the polarized target asymmetry, when compared to the data, gives confidence
to the presence of a third $S_{11}$ resonance, for which we extract some static and
dynamical properties: $M \approx$ 1.730 GeV, $\Gamma_{T} \approx$ 180 MeV.
These results are in very good agreement with those in Ref. \cite{zr96}, and compatible
with ones in Ref.~\cite{Sva2}.

The dynamics of our models is partially based on the duality hypothesis, namely,
the exclusion of the $\rho$ and $\omega$ vector mesons exchange in the {\it t}-channel.
However, our approach allows us to take into account individual contributions
from all known nucleon resonances up to $F_{15}$(2000), 
and treat as degenerate higher ones up to $G_{17}$(2190). 
These facts seem to us 
reasonable justification to apply that hypothesis. Actually, the
manifestations of the duality in the case of pseudoscalar mesons have
been discussed in detail in Ref.~\cite{ST}. In this latter study, it
was shown that the {\it t}-channel exchanges mimick the {\it higher spin}
resonances lacking in the models. 
In the present work, given the kinematics
region under consideration, we do not expect significant contributions from
resonances with spin and mass higher than those of the $G_{17}$(2190) 
resonance. 
Moreover,
the new resonance comes out to be an $S_{11}$-wave, while the contributions
from higher spin and mass resonances, $P_{13}$(1900) and $F_{15}$(2000),
are found marginal. 
More generally,
in searching for new resonances, it is highly desirable to avoid
{\it t}-channel contributions in order not to wash out possible 
manifestations of yet undiscovered resonances.

If the trend seen in the data from Graal is confirmed by higher energy 
measurements, the existence of the new $S_{11}$ resonance could be endorsed
and this finding will have
very important implications. Actually, this new resonance can not be accommodated by 
the constituent quark
model, which may suggests an exotic nature for this resonance, such as 
a $\Sigma K$ or $\Lambda K$ type molecule~\cite{zr96}. 
If this is indeed the case, the investigation of this resonance 
in other reactions might be certainly warranted to understand 
its internal structure.  For example, a systematic study
of the kaon photoproduction in the threshold region in different isospin channels would 
certainly shed some light on this resonance. The $\eta $ and kaon electroproduction will also
be very desirable to study
the $Q^2$ dependence of transition amplitudes. The QCD counting rule implies  
different $Q^2$ dependence for resonances with a three quark constituent or 
other types of internal structure. 

Certainly, forthcoming data from existing facilities will
provide us with more information on the existence and the nature of this resonance.

%%%%%%

\section{Acknowledgments}
\label{acknowledgements}
We wish to thank the GRAAL collaboration and especially,
J.-P. Bocquet, D. Rebreyend, and F. Renard for 
having provided us with their data prior to publication, 
and fruitful exchanges. 
We are also grateful to S. Dytman and B. Ritchie for
enlightening discussions on the measurements performed
at JLab.

%
% Non-BibTeX users please use

\end{document}